\documentclass[conference]{IEEEtran}
\IEEEoverridecommandlockouts
\usepackage{etex}
\usepackage{booktabs}
\usepackage[utf8]{inputenc}
\usepackage[ruled,vlined]{algorithm2e}
\usepackage[T1]{fontenc}
\usepackage{microtype}
\usepackage{graphicx}
\usepackage{paralist}
\usepackage{tabularx}
\usepackage{balance}
\usepackage{multicol}
\usepackage{multirow}
\usepackage{pbox}
\usepackage{enumitem}
\usepackage{amssymb}
\usepackage{lscape}
\usepackage{pifont}
\usepackage{xspace}
\usepackage{url}
\usepackage{tikz}
\usepackage{rotating}
\usepackage{float}
\usepackage[TABBOTCAP]{subfigure}
\usepackage{ragged2e}
\usepackage{fontawesome}
\usepackage[colorlinks=true, allcolors=blue]{hyperref}

\usepackage{cite}
\usepackage{amsmath,amssymb,amsfonts}
\usepackage{algorithmic}
\usepackage{graphicx}
\usepackage{textcomp}
\usepackage{xcolor}
\def\BibTeX{{\rm B\kern-.05em{\sc i\kern-.025em b}\kern-.08em
    T\kern-.1667em\lower.7ex\hbox{E}\kern-.125emX}}

  {\list{}{\leftmargin=0.1in\rightmargin=0.1in}\item[]}%
  {\endlist}

\newcommand{\ie}{\emph{i.e.,}\xspace}
\newcommand{\eg}{\emph{e.g.,}\xspace}
\newcommand{\etc}{etc.\xspace}
\newcommand{\etal}{\emph{et~al.}\xspace}
\newcommand{\secref}[1]{Section~\ref{#1}\xspace}

\newcommand{\figref}[1]{Fig.~\ref{#1}\xspace}

\newcommand{\tabref}[1]{Table~\ref{#1}\xspace}

\newcommand{\totalDevs}{80\xspace}
\newcommand{\devFirstStudy}{80\xspace}
\newcommand{\characteristics}{70\xspace}

\newcommand*\circled[1]{\tikz[baseline=(char.base)]{
    \node[shape=circle,draw,inner sep=1pt] (char) {#1};}}

\begin{document}

\pagestyle{empty} 

\title{Source Code Recommender Systems:\\The Practitioners' Perspective}

\author{
\IEEEauthorblockN{Matteo Ciniselli\IEEEauthorrefmark{1}, Luca Pascarella\IEEEauthorrefmark{1}, Emad Aghajani\IEEEauthorrefmark{1}, Simone Scalabrino\IEEEauthorrefmark{2}, Rocco Oliveto\IEEEauthorrefmark{2}, Gabriele Bavota\IEEEauthorrefmark{1}}
\IEEEauthorblockA{\IEEEauthorrefmark{1}\textit{SEART @ Software Institute, Universit\`{a} della Svizzera italiana (USI), Switzerland}}
\IEEEauthorblockA{\IEEEauthorrefmark{2}\textit{STAKE Lab @ University of Molise, Italy}}
}

\maketitle

\begin{abstract}
\justifying
The automatic generation of source code is one of the long-lasting dreams in software engineering research. Several techniques have been proposed to speed up the writing of new code. For example, code completion techniques can recommend to developers the next few tokens they are likely to type, while retrieval-based approaches can suggest code snippets relevant for the task at hand. Also, deep learning has been used to automatically generate code statements starting from a natural language description. While research in this field is very active, there is no study investigating what the users of code recommender systems (\ie software practitioners) actually need from these tools. We present a study involving \totalDevs software developers to investigate the characteristics of code recommender systems they consider important. The output of our study is a taxonomy of \characteristics ``requirements'' that should be considered when designing code recommender systems. For example, developers would like the recommended code to use the same coding style of the code under development. Also, code recommenders being ``aware'' of the developers' knowledge (\eg what are the framework/libraries they already used in the past) and able to customize the recommendations based on this knowledge would be appreciated by practitioners. The taxonomy output of our study points to a wide set of future research directions for code recommenders.
\end{abstract}


\IEEEdisplaynontitleabstractindextext
\IEEEpeerreviewmaketitle

\section{Introduction} \label{sec:intro}

Recommender systems are becoming more and more popular in software engineering. These tools can support developers in several tasks \cite{robillard:recommenders}, such as documentation writing and retrieval \cite{Xie:msr2006,Moreno:icse2015,Moreno:tse2016,Xing:icpc2018}, code refactoring \cite{Bavota:emse2014,Tsantalis:saner2018}, bug fixing \cite{goues:icse2012,Tufano:tosem2019,Li:icse2020}, bug triaging \cite{Tamrawi:2011,Xia:tse2016}, code review \cite{Tufano:icse2021,Tufano:icse2022} \etc Among these, \emph{source code recommender systems} support developers in writing code. 

Code recommender systems have been designed using different underlying techniques. For example, retrieval-based approaches \cite{McMillan:tse2013,Wen:icse2021} identify relevant code elements to reuse given the code under development in the Integrated Development Environment (IDE) and/or a query describing the coding task at hand. Other approaches exploit deep learning (DL) to (i) automatically generate a needed code (or parts of it) given its textual description \cite{Nguyen:fse2016,xu-etal-2020-incorporating} or (ii) predict the tokens the developer is likely to type given the code in the IDE~\cite{Liu:ase2020,wang2020towards}. Building on top of this literature, GitHub recently presented Copilot \cite{copilot}, a tool using DL to recommend entire code statements or even whole functions. \eject

While research on code recommender systems is extremely active, no previous work investigated what the \emph{desiderata} of software developers are. In other words, the techniques proposed in the literature are mostly based on assumptions made by researchers. For example, a recent work by Wen \etal \cite{Wen:icse2021} targets the automatic implementation of whole code functions. However, it is unclear whether developers are actually looking for such a type of support or if, instead, they prefer the classic code completion implemented in IDEs. Also, none of the existing tools and techniques customize their code recommendations based on the developer's coding style and/or to their expertise and it is unclear whether such a functionality would be important for practitioners. To answer these questions, we present a study involving \totalDevs practitioners which is aimed at investigating the characteristics of code recommenders that they consider important. In particular, after collecting demographics information, we asked participants their opinion about the code recommenders they use (\eg copilot, default code completion in IDE, \etc) from three perspectives: (i) \emph{coverage} (\ie in how many coding scenarios the tool can provide recommendations), (ii) \emph{accuracy} (\ie to what extent recommendations are close to what practitioners need), and (iii) \emph{usability} (\ie how friendly the user interface is). For each of these three aspects participants were asked to describe improvements (if any) they would like to see in the recommender they mostly use. Then, we ask them in an open-ended question what the characteristics of code recommenders they consider important are. 

Each answer we received has been independently analyzed by five authors through an open-coding inspired approach with the goal of assigning a set of tags to it. The extracted tags represent requirements expressed by practitioners for code recommenders and, after conflict resolution, have been organized in a hierarchical taxonomy. Such a process has been performed iteratively four times, with 20 practitioners taking part in each iteration. We stopped once the output taxonomy converged (\ie no additional requirements were added to the taxonomy from the answers we received in the last iteration). 

The output of our study is a taxonomy of \characteristics ``requirements'' that should be considered when designing code recommender systems (\figref{fig:results}). For example, our taxonomy highlights that practitioners are  interested in \emph{adaptive} recommendations, meaning that the recommended code should be automatically adapted to the code under development (\eg reusing identifiers when possible) and to the developer's coding style. \smallskip

\textbf{Significance of research contribution.} The taxonomy of \characteristics ``requirements'' for code recommenders output of our study provides a rich research roadmap in the field of code recommender systems. Indeed, as our empirical evidence shows, the \emph{desiderata} of practitioners in this context are not always aligned with what offered by state-of-the-art techniques. 

Based on our taxonomy, researchers can have a clear understanding of what the priorities are when designing code recommenders. 

\textbf{Data availability.} We release the survey used in the study, the collected answers with the results of the manual analysis we performed on them, plus additional material in our replication package: \underline{\url{https://code-recommenders.github.io}}.

\section{Study Design} \label{sec:study1}

The \emph{goal} of the study is to investigate what the \emph{desiderata} of software practitioners are when it comes to code recommender systems. The \emph{context} consists of \emph{objects}, \ie a survey designed to investigate the study goal, and \emph{subjects} (referred to as ``participants''), \ie \devFirstStudy practitioners recruited through Amazon Mechanical Turk (MTurk) \cite{mturk} and personal contacts. \smallskip

We aim at answering the following research question:

\begin{quote}

{\em What are the characteristics of code recommender systems that are considered important by practitioners?} Despite the many code recommender systems proposed in the literature, no study investigated what practitioners actually need in terms of automatic support during coding activities. Answering our RQ can guide the development of better code recommender systems.

\end{quote}

\subsection{Context Selection --- Participants} \label{sec:study1participants}
We recruited participants through two channels. First, we used MTurk \cite{mturk}, a crowdsourcing website to hire people for on-demand tasks. We enrolled participants that (i) have successfully completed in the past at least 50 tasks on MTurk; (ii) have an approval rate for their past tasks grater than 90\% (\ie more than 90\% of the tasks they performed in the past have been approved by the requester); (iii) hold the MTurk Master qualification assigned to ``top workers''. We only involved practitioners in our survey, excluding students (at any level) and researchers.

Participants who completed our survey were payed 10\$ upon a manual verification in which we made sure that the provided open answers (described in the following) were meaningful and written in correct English. We collected 31 complete surveys from MTurk. In addition, we invited practitioners in the authors' contact network. This resulted in additional 49 answers, leading to a total of \devFirstStudy participants. As explained later, developers were not invited all together, but in four rounds of 20 participants each. Demographics about participants are presented in \secref{sec:results}.

\subsection{Context Selection --- Survey} \label{sub:survey1}
Our survey has been implemented in Qualtrics \cite{qualtrics} and is available in our replication package \cite{replication}. 

Participants were initially presented with a welcome page, which explained the goal of the study, reported its expected duration ($\sim$15 minutes), and set the context by explaining what source code recommender systems are: 

\begin{quote}
\emph{With source code recommender systems we refer to approaches that can be used to automatically suggest code to developers while they are writing code. The classic example in this context is the code completion feature implemented in IDEs.\\However, some tools go beyond the classic code completion task and recommend longer pieces of code to developers to autocomplete a task they perform (see, \eg \url{https://copilot.github.com/}).}
\end{quote}

By agreeing to participate, they started our survey composed of three steps.\smallskip

\textbf{Step \circled{1}: Demographic Information}. We asked participants to indicate (i) their job position (\eg developer, tester), (ii) the programming language and the IDE they mostly use, and (iii) the number of years of programming experience.\smallskip

\textbf{Step \circled{2}: Experience with Code Recommenders}. The second step included questions about the participants' experience with code recommender systems. We asked what tool(s) they use as code recommender, with the possibility of selecting ``\emph{The default one in the IDE}'' and/or specifying the tool(s) in an open text box. If participants indicated that they did not use any code recommender, the survey stopped. We also collected the frequency with which participants check code recommendations: Occasionally (\ie less than 50\% of times a recommendation is available), Most of times (\ie more than 50\%, but not always), and Always. 

\begin{table*}[]
\centering
\caption{Characteristics of code recommendations collected from literature}
\begin{tabular}{lm{12.5cm}c}\toprule
\textbf{Attribute}         & \textbf{Description} & \textbf{References} \\ \midrule
Concise Code               & The recommended code must be as short and simple as possible.                                                                                                                                                                 & \cite{Kim:ASE2009,Nasehi:icsm2012,hellendoorn2019code}                    \\ \midrule
Correct Code               & The recommended code must be bug-free.                                                                                                                                                                                        &         \cite{Kim:ASE2009}            \\ \midrule
Familiar                   & If multiple recommendations are possible, the one using code that is more familiar to the developer must be used (\eg the code using APIs already used in the past by the developer receiving the recommendation).          &         \cite{marri2009improving}            \\ \midrule
High readability           & The recommended code must be readable (\eg avoid very long statements, adopt indentation).                                                                                                                                  &  \cite{Moreno:icse2015}                   \\ \midrule
High reusability           & The recommended code must be easy to reuse (\eg a code using object types not available in the language but defined in other projects from which the recommended code has been learned is difficult to reuse).              &   \cite{Moreno:icse2015}                  \\ \midrule
Inline Documentation       & The recommended code features comments explaining the code step-by-step.                                                                                                                                                        &  \cite{Nasehi:icsm2012}                   \\ \midrule
Meets coding layout        & The recommended code must be adapted to the context of the recommendation by adopting the same coding layout (\eg same indentation, spaces between code tokens).                                                            &            \cite{kyaw2018proposal}         \\ \midrule
Meets naming conventions   & The recommended code must be adapted to the context of the recommendation by adopting the same naming conventions (\eg if variables are named with camelCase, the same convention must be adopted in the recommended code). &        \cite{kyaw2018proposal}             \\ \midrule
Precise Typing Information & A recommended code using AVerySpecificType should be preferred over a recommended code using Object.                                                                                                                          &     \cite{Perelman:pldi2012}                \\ \midrule
Responsive                 & The responsiveness of the code recommendation system in terms of time needed to generate a recommendation.                                                                                                                    &             \cite{svyatkovskiy2020intellicode}        \\ \midrule
Same name                  & The recommended code must be adapted to the context of the recommendation by using the same variable names of the code it completes when possible.                                                                            &       \cite{Perelman:pldi2012}              \\ \midrule
Syntactical Correctness    & The recommended must not introduce syntax errors.                                                                                                                                                                             &       \cite{Amorim:sle2016,wang2020towards,svyatkovskiy2020intellicode}              \\ \midrule
Step-by-step Solution      & In case the recommended code spans across many statements, the code is divided into multiple chunks (by using a blank line), each one responsible for a sub-task.                                                             & \cite{Nasehi:icsm2012}                    \\ \midrule

Vulnerability-free         & The recommended code must be vulnerability-free.                                                                                                                                                                              &     \cite{schuster2020autocomplete}                \\\bottomrule                  
\end{tabular}
\label{tab:literatureChar}
\vspace{-0.2cm}
\end{table*}

Then, we asked to rate the code recommendation capabilities of the tool they mostly use from three perspectives: (i) \emph{coverage} (\ie in how many coding scenarios the tool can provide recommendations), (ii) \emph{accuracy} (\ie to what extent recommendations are close to what practitioners need), and (iii) \emph{usability} (\ie how friendly the user interface is). For each of these three aspects, participants could indicate their answer on a three-point scale (Low, Medium, High), or select a ``Not sure'' option. We provided participants with detailed explanations about the three perspectives \cite{replication}. For each of them, we also asked the improvements participants would like to see in the code recommender(s) they commonly use (\eg what they would like to have in terms of ``coverage'' that is not currently supported). Finally, we ask a specific question related to the accuracy of the recommendations: \emph{Assume that a code recommendation tool provides a list of recommendations sorted by likelihood of being relevant (\ie the most relevant on top). How many of these recommendations would you be willing to read to find the right one?} Answers to this question can inform the evaluation of code recommenders by researchers. Indeed, when evaluating approaches for automatically generating code, researchers often assess their performance for the top-$k$ recommendations (\eg top-50 in \cite{Tufano:tosem2019}). Knowing how many recommendations developers are willing to inspect can help in setting the number of generated solutions to realistic values. \smallskip

\textbf{Step \circled{3}: Characteristics of Code Recommenders}. The third and last step is the core of our survey, in which we asked participants the characteristics of code recommendations they consider important. In the first question, participants could describe in an open text box the characteristics of code recommendations they perceive as most important, accompanying each one with a short explanation/rationale. We clarified that they could include both functional and non-functional characteristics. 

In the second question, the survey showed a list of 14 characteristics we preliminarily defined, asking participants to select the ones they consider important (if any). To determine such 14 categories, we manually analyzed the state of the art related to code recommender systems. Specifically, we focused on papers that presented: (i) techniques for code generation/completion (see \eg \cite{svyatkovskiy2020intellicode}), (ii) empirical studies about code generation/completion techniques (\eg \cite{schuster2020autocomplete}), (iii) techniques to generate code examples (\eg \cite{Moreno:icse2015}), and (iv) empirical studies about code examples used by developers (\eg \cite{Nasehi:icsm2012}). For example, we extracted the \emph{high reusability} characteristic from the paper by Moreno \etal \cite{Moreno:icse2015}.  

The full list of characteristics with the corresponding references they were extracted from is available in \tabref{tab:literatureChar}. Note that, to define such a list, we did not perform a systematic literature review to identify \textbf{all} papers in the surveyed areas: We relied on the experience of the six authors to identify a set of 41 peer-reviewed papers published in international conferences and journals and applied backward snowballing on their references to identify additional relevant works. 

At the end, we inspected 53 papers. Each paper was assigned to one author in charge of adding to a spreadsheet the list of ``characteristics'' described in the paper (if any). In some cases, the papers from which we extracted a characteristic did not explicitly point to the need for considering such a characteristic when building code recommender systems. 

However, this could be inferred from the text of the paper. One example is the work by Schuster \etal \cite{schuster2020autocomplete}. The authors show that ``\emph{neural code autocompleters are vulnerable to poisoning attack}'' \cite{schuster2020autocomplete}. Thus, we infer that \emph{Vulnerability-free} is one of the characteristics to assess for code recommenders. Similarly, Nasehi \etal \cite{Nasehi:icsm2012} studied what makes a good code example on Stack Overflow. Again, this is something not directly linked to a code recommender. However, we assume that characteristics of good code examples could be relevant for recommended code as well. 

\begin{table}[ht]
	\centering
	\caption{Rounds of data collection (20 participants each)}
	\label{tab:rounds}
	\begin{tabular}{lrrrrr}
	\toprule
	\multirow{2}{*}{\textbf{Round}} & \textbf{New L1} & \textbf{New L2} & \textbf{New L3} & \textbf{New L4} & \multirow{2}{*}{\textbf{Conflicts}}\\
	& \textbf{Charac.} & \textbf{Charac.} & \textbf{Charac.} & \textbf{Charac.} & \\\midrule
	I & 5 & 25 & 13 & 0 & 24\%\\
	II & 0 & 9 & 6 & 6 & 21\%\\
	III & 0 & 0 & 6 & 0 & 24\%\\
	IV & 0 & 0 & 0 & 0 & 24\%\\
	\bottomrule
\end{tabular}

\end{table}

\eject

\subsection{Data Collection and Analysis}
Given the goal of our study, we decided to run it in multiple rounds until we reached saturation in the collected taxonomy of characteristics. First, we invited 20 practitioners to complete our survey (first round). Then, to answer our RQ, we analyzed the data collected in steps \circled{2} and \circled{3} in the open answers by following an open-coding inspired approach: Five of the authors independently assigned a set of tags to each of the open answers provided by the 20 participants that described through written text the improvements they would like to see in terms of coverage, accuracy, and usability of the code recommenders they use and the characteristics of code recommenders they perceive as most important. Each tag was meant to encode a specific characteristic (\eg \emph{Early prediction} was derived from the answer ``\emph{if the completions are not timely, it is just easier to type the whole thing myself sometimes}''). 

Conflicts (\ie different tags assigned by the five authors to the same answer) were solved through online meetings involving all authors. The set of ``characteristics'' derived through this analysis was then complemented with the ones selected by developers as important from the list extracted from the literature. The output of this analysis is a hierarchical taxonomy of characteristics of code recommenders (\figref{fig:results}), with level-1 nodes indicating root categories, level-2 nodes indicating sub-categories of a specific root category, and so on.

This first round was followed by three additional rounds, each of which added 20 participants. We stopped with this process when the execution of a new round did not result in the addition of any new characteristic in our taxonomy, indicating that a good level of saturation was reached. We are aware that additional rounds may further strengthen the generalizability of our taxonomy. However, we had to balance the comprehensiveness of the taxonomy with the feasibility of the study given our limited access to professional developers. 

\tabref{tab:rounds} reports for each round (i) the number of new characteristics added to the different levels of the taxonomy and (ii) the percentage of conflicts arisen during open coding. The number of characteristics reported in \tabref{tab:rounds} for each level does not match the final number in \figref{fig:results}: This happens because we reorganized the taxonomy after each round for readability reasons (\eg some level-2 characteristics were moved to level-3 as child of a new level-2 category).
However, the overall number of characteristics (\characteristics) matches the one in our final taxonomy (\figref{fig:results}).

Concerning conflicts, since we had five authors inspecting each answer in an open coding setting in which the codes (\ie characteristics of code recommenders) to extract were not pre-defined but had to emerge from the data, we had some form of conflict very frequently, \eg one of the five authors not extracting a characteristic indicated by the remaining four authors. These cases were usually trivial to solve in the online meetings. Other types of conflict required instead longer discussions, and these are the ones we document in \tabref{tab:rounds}. 

\eject

In particular, we report the percentage of cases in which there was no majority in extracting a ``characteristic'' from an open answer (\ie less than three authors reported it). As it can be seen, this happened in $\sim$20\% of cases in each round.


\begin{figure}[ht]
\centering
\includegraphics[width=\linewidth]{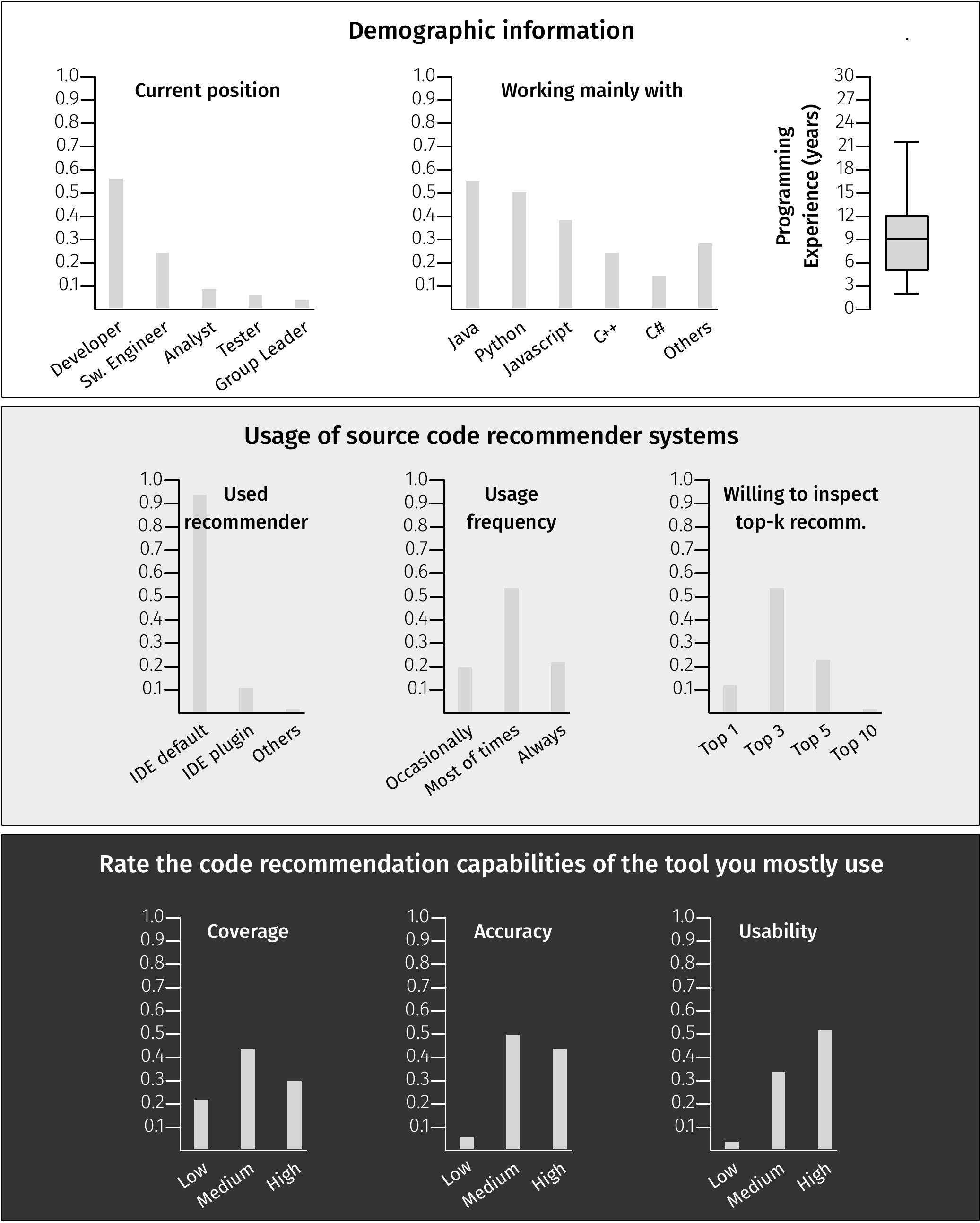}
\caption{Demographic information (top), usage of source code recommenders (middle), and assessment of code recommender systems (bottom).}
\label{fig:demographics}
\end{figure}

\section{Results Discussion} \label{sec:results}
We start by summarizing demographic information about the study participants and their experience with code recommender systems. Then, we discuss the taxonomy of characteristics of code recommender systems, which is the main outcome of this study.

\subsection{Demographics and Experience with Code Recommenders}
\figref{fig:demographics} presents a visualization of demographic data for the \devFirstStudy practitioners involved in our study. Most of the participants are developers (47) or software engineers (20), with others classifying themselves as analysts, testers, and group leaders. Participants mostly work with Java, Python and/or Javascript. 

In terms of experience, 86\% of participants has more than five years of programming experience, with an average of 9.7 years, and a median of 9. The vast majority (85\%) of participants use the IDE's built-in code completion feature as their \emph{only} code recommender system. 

\eject

Among the used IDE plugins, GitHub Copilot is the most represented one, with 5 mentions all coming from the latest round we performed. Indeed, when we run the first three rounds Copilot was not yet publicly available. 

Less than 19\% of the participants claimed to ``occasionally inspect the code recommendations provided'', with the remaining ones looking at them \textit{most of times} or \textit{always} when they are available. 

\begin{figure*}[tb]
	\centering
	\includegraphics[width=1\linewidth]{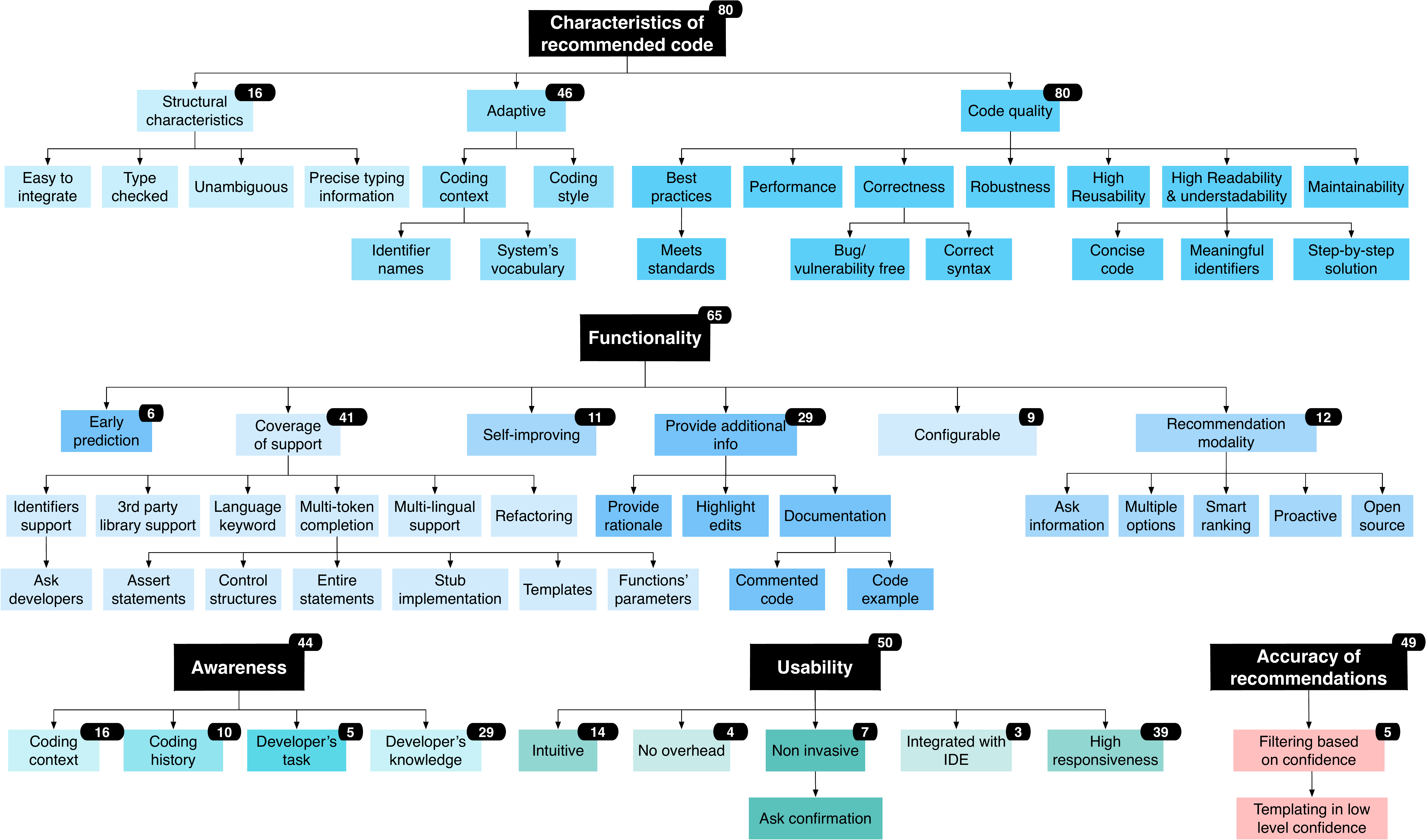}
	\caption{Taxonomy of characteristics of code recommender systems}
	\label{fig:results}
\end{figure*}

\begin{figure*}[h!]
	\centering
	\includegraphics[width=\linewidth]{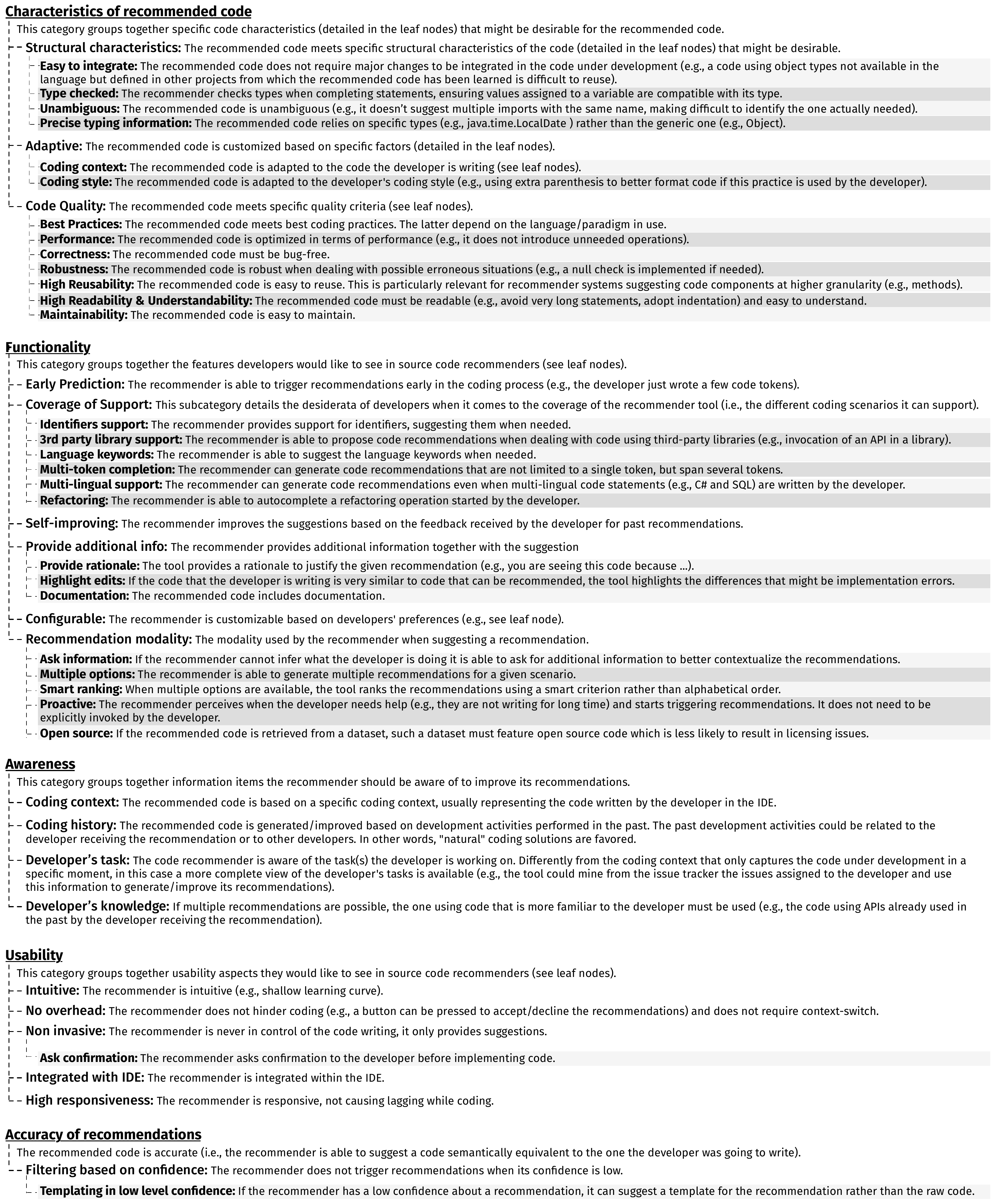}
	\caption{Definitions for taxonomy's characteristics up to level-3}
	\label{fig:definitions}
\end{figure*}

Participants indicated the willingness to inspect at most the top-5 code recommendations provided (95\%), with the majority (56\%) focusing only on the top-3. This suggests researchers to limit the assessment of code recommenders to the top-5 recommended solutions, since others are very unlikely to be considered by developers.

When asked about their assessment of the code recommender they use in terms of coverage, accuracy, and usability, developers reported that they are mostly happy of the tools they use as for these aspects. Specifically, $\sim$44\% of them considered the accuracy high, and $\sim$50\% medium; $\sim$58\% evaluated the usability as high, and $\sim$38\% as medium; and $\sim$31\% judged the coverage high and $\sim$46\% medium. The coverage of the provided support (\ie the variety of scenarios in which the tool is able to recommend code) is the aspect achieving the lowest rates, with $\sim$23\% of participants reporting a low coverage.

\subsection{Taxonomy Discussion}

\figref{fig:results} reports the taxonomy of characteristics participants indicated as relevant for code recommender systems. The number on the top-right of the level-1 and level-2 categories indicates the number of participants who mentioned such a characteristic as important (out of \devFirstStudy). While some characteristics have only been mentioned by few developers (\eg \emph{Awareness} $\rightarrow$ \emph{Developer's task}) we decided to include all of them for completeness. \smallskip

On top of that, \figref{fig:definitions} reports the definition of all categories up to level-3. We do not report the ones for level-4 categories for space reason. However, these usually represent specifications of level-3 categories that are quite intuitive and we include in our replication package \cite{replication} complete definitions for all categories, with a search interface that helps in quickly identify the category of interest. 

We discuss our taxonomy, going through each of the five root categories depicted in \figref{fig:results}. We use the \faLightbulbO~icon to highlight lessons learned for future research in the field. \smallskip

\textbf{Characteristics of recommended code.} 
This root category groups characteristics of the recommended code that participants consider important. All participants mentioned at least one aspect related to the \emph{quality} of the recommended code. Note that, at a first sight, one may think that code quality is not relevant for code recommenders, since they often recommend a few code tokens to complete a statement the developer is writing. This is, however, not the case for the new generation of code recommenders (see \eg GitHub Copilot \cite{copilot} or recently proposed works in the literature \cite{svyatkovskiy2020intellicode,Wen:icse2021,ciniselli:tse2021}) that can recommend complete code statements or even entire functions. 

Receiving recommendations having a high readability and understandability is a priority for developers (66 mentions). 39\% of the participants indicated their preference for concise recommendations, \eg ``\emph{I do not use completion that suggests entire snippets. The reason is that if I get an entire snippet I'll need to understand it to make sure is what I need and, based on my experience, this is not faster than writing the code myself}.'' \faLightbulbO~This goes somehow in contrast with recent work targeting the automatic implementation of whole functions \cite{Wen:icse2021,copilot}. However, as we will see later when discussing the \emph{Coverage of Support}, developers are willing to use recommendations for complex scenarios (such as entire functions) if they have a high confidence in the received recommendations. Also, in case of recommendations composed by several statements, 14 participants indicated the importance of organizing these recommendations as \emph{Step-by-step solutions}, meaning that the code is divided into multiple chunks (using a blank line), each one responsible for a sub-task. Such a feature requires the ability of the recommender to identify sub-tasks within the suggested code.

\faLightbulbO~More in general, readability and understandability are two important aspects of code recommendations, \eg ``\emph{If I don't understand what the suggestion is about at a first glance, I ignore it and I continue programming}''. However, to the best of our knowledge, no code recommender explicitly focuses on these aspects when deciding which recommendation to trigger. While this could be possible exploiting the readability metrics previously defined in the literature \cite{Buse:tse2010, posnett2011simpler, dorn2012general, scalabrino2018comprehensive, mi2018improving}, it is still unclear to what extent such metrics work on artificial code. 

Other participants pointed at the importance of the recommended code to meet \emph{best practices} and \emph{coding standards}. \faLightbulbO~This includes the possibility to customize the notion of coding standard (\eg ``\emph{meets coding standards of the company in terms of code quality}'', ``\emph{pushing good coding standards, either general ones or customized by the user}''). This is another aspect currently unsupported in the state of the art.

Several other aspects of code quality have been mentioned by participants (\eg ensure good \emph{performance}, \emph{robustness}, and \emph{reusability} of the recommended code). For example, in the case of performance one practitioner wrote: ``\emph{Now that more complex recommendations are possible thanks to tools like copilot, aspects related to code quality should be taken into account more, for example by picking among two possible recommendations the one ensuring better performance}''. 

A crucial quality aspect mentioned by 69 participants is, as expected, the \emph{correctness} of the recommended code. This means that the recommendation must not break the syntax and/or introduce bugs/vulnerabilities in the code under development (\eg ``\emph{the IDE must never recommend a solution that leads to a malfunction}''). While such a finding might look obvious, it triggers a few considerations about state-of-the-art techniques. For example, DL models have been proposed to support code completion \cite{ciniselli:tse2021,Liu:ase2020,wang2020towards}. 

\eject

However, recent work by Ciniselli \etal \cite{ciniselli:tse2021} showed that, in the best case scenario, these models can recommend a correct code completion in less than 70\% of cases, with much worst results ($<$30\%) achieved for challenging completions (\eg recommend entire statements). While the study by Ciniselli \etal assessed the accuracy of the recommendations (\ie the extent to which the tool completes the code as expected), it is likely that a tool not providing an accurate recommendation breaks the code syntax or even introduces bugs. \faLightbulbO~Our survey suggests that developers are unlikely to use tools that could potentially ``break'' their code in $\sim$35\% of cases. For the reasons discussed, the notion of correctness is strongly related to the \textbf{accuracy of recommendations}, that is another root category in our taxonomy (see \figref{fig:results}). 

The majority of surveyed developers (61\%) reported the accuracy of recommendations as an important characteristic of code recommenders (\eg ``\emph{inaccurate recommendations would create more work than manually writing the code}'', ``\emph{the recommender system should have a minimum amount of false positive hits}'', ``\emph{accuracy below 80\% is not acceptable}''). \faLightbulbO~Some developers even proposed solutions to overcome issues related to the low accuracy in specific scenarios (these led to the two child categories of \emph{Accuracy of recommendations} in \figref{fig:results}). For example, a participant mentioned ``\emph{it must be possible to configure the suggestions, to get less but more likely to be correct}''. 

\eject

Basically, the possibility to filter out recommendations in which the recommender system has a low confidence could help in addressing limited accuracy in challenging scenarios. Some developers also indicated their willingness to consider recommendations that are not 100\% accurate as long as they can be easily modified to obtain the needed code: ``\emph{I often accept the recommendation even if it's not 100\% accurate if I think that adjusting it takes less time than writing code from scratch}''.

Going back to the \emph{characteristics of recommended code} tree, developers are also looking for \emph{adaptive} recommender systems: Developers mentioned that the tool should adapt the recommendations to their coding style (\eg ``\emph{once it learns my coding style it should go with that}'') and to the coding context (\eg ``\emph{context-sensitivity is also important; I would like practical and stylistic compatibility with the existing code}''). 

\faLightbulbO~Automatically inferring the developer's coding style is far from trivial. One possibility would be to investigate the integration between approaches to learn coding style/conventions \cite{Reiss:ase2007,Allamanis:fse2014} and code recommender systems. More in general, our survey seems to suggest the need for ``modeling software developers and their coding practices'' with all difficulties and privacy concerns this may lead to. Having such information may substantially boost the usefulness of code recommenders, that could better adapt their recommendation to the user receiving them. 

\eject

Finally, the \emph{structural characteristics} subtree groups characteristics of the recommended code that concern its structure. These have been considered relevant by a lower number of developers (16). An interesting point to highlight here is the need for code recommendations that are \emph{easy to integrate} in the code under development. This aspect is related to the adaptability of the recommendations: if the recommended code adapts to the coding context, for example, reusing identifiers when needed, it can make the developer's life easier. This is in contrast with what has been done in retrieval-based approaches that recommend relevant pieces of code from a code base (\eg Stack Overflow) leaving the integration effort on the developer's shoulders. \faLightbulbO~An approach able to automate, at least in part, the integration process could substantially increase the usefulness of code recommenders.\smallskip
 
\textbf{Functionality.}
This category groups features developers would like to see in code recommenders. Most of the answers point to specific wishes in terms of \emph{coverage of support}. This means that developers would like to have a wider support in terms of code recommendations, something that goes beyond the tools they currently use. Such a subtree supports several of the directions that the software engineering research community is currently investigating.

Approaches for \emph{multi-token completion} (\ie code completion that goes beyond recommending the next token the developer is likely to write, for example, an entire statement) are a hot research topic \cite{kim2020code,Karampatsis:DLareBest,Liu:ase2020,ciniselli:tse2021,copilot} as well as the automated generation of \emph{assert statements} \cite{Watson:icse2020,Tufano:asserts} (see \figref{fig:results}). \faLightbulbO~Our results show the need for techniques able to support developers in complex code completion scenarios (\eg ``\emph{code completion provides good support in finding the right API in a class and a few other things, but rarely suggests something challenging such as conditional statements}'', ``\emph{cool would be suggesting which asserts are needed to test the code under development}''). While no developer asked for the automated implementation of whole functions from scratch \cite{Wen:icse2021}, some mentioned the possibility to automatically implement \emph{stub functions} starting from their signature as recently done in Copilot \cite{copilot}: \faLightbulbO~``\emph{When writing new functions I often create stubs of the other functions I need to invoke; the IDE could propose implementations for those stubs}''.

\faLightbulbO~An interesting research challenge also comes from the suggestion for \emph{multi-lingual support}: ``\emph{autocomplete when mixing languages such as inline SQL code when writing database statements in C\#}''. Such a support is currently missing in code recommenders. Finally, for what concerns the \emph{coverage of support}, it is also interesting the possibility to integrate the capabilities to autocomplete \emph{refactoring} operations started by the developer. This is something that has been accomplished by Foster \etal \cite{Foster:icse2012} with their WitchDoctor tool. 

Another representative category in the functionality tree is ``\emph{provide additional info}'', which relates the will of developers to receive additional information accompanying the code recommendation. 

\eject

This includes the possibility to \emph{provide a rationale} justifying non-trivial code recommendations (\eg ``\emph{it should give a quick reason that I can understand why it's suggesting I do it}''), or documenting the recommended code (\eg ``\emph{assuming the recommended code is not trivial, a documentation/explanation for the suggestion is needed}''). 

While code recommenders mostly focus on generating meaningful code recommendations, retrieval-based techniques (\ie those retrieving relevant code from a code base) and properly trained DL-based techniques can provide support for the automated documentation of the recommended code \cite{huang2020towards,copilot}. \faLightbulbO~More challenging is the generation of a rationale explaining to the developer why a given recommendation is relevant for what they are doing.

Finally, the other child categories under \emph{functionality} have been mentioned by a few developers. They concern: (i) the ability of the tool to improve over time based on accepted/rejected recommendations (\emph{self-improving}); (ii) the possibility to configure aspects of the tool, such as defining shortcuts (\emph{configurable}); (iii) the way in which the recommendations are presented or generated (\emph{recommendation modality}), and (iv) the need for having the recommendation quickly triggered to avoid manually writing most of the code thus reducing the usefulness of the recommendation (\emph{early prediction}). For the sake of brevity, these categories are described in \figref{fig:definitions}.\smallskip

\textbf{Usability.}
This root category concerns aspects that influence the usability of the code recommender. Among all, \emph{high responsiveness} was the most important requirement highlighted by developers (39 mentions) (\eg \emph{``these recommendation systems are often slow and lag the entire UI, especially on large projects''}). \faLightbulbO~This confirms the relevance of works investigating efficiency aspects in code completion tools \cite{svyatkovskiy2020fast}.

Along this line, participants expect tools which are \emph{intuitive} (\eg allowing a gradual learning curve for all developers) and \emph{well integrated with IDE}, as one developer underlined: \emph{``clean interface making it easy to use is pretty important''}.

Finally, an interesting aspect highlighted by some developers is the need for \emph{non invasive} code recommenders, always leaving the final word to the developer: ``\emph{It shouldn't change the code without my confirmation, even though the code I'm writing is not logically right; a warning would be nice}''. \smallskip

\textbf{Awareness.}
The last root category in our taxonomy is \emph{awareness}: the code recommender must be ``aware'' of different aspects to improve its recommendations. Most importantly from the participants' point of view (29 mentions) is the awareness of developer's knowledge. This implies that the same recommender system triggered on the same code by two different developers could produce different recommendations. \faLightbulbO~Code recommendations using APIs familiar to the developer (\eg that the developer used in the past) can be favored as well as recommendations using code constructs that are familiar to the developer. As previously observed, this requires the ability of the recommender to ``model'' the developer's knowledge, exploiting it in generating the recommendations with the goal of minimizing the comprehension effort for the developer.

A second important aspect is the awareness of \emph{coding context}, \ie the code recommender should consider the current code context (\eg the code the developer is writing in the IDE) when providing suggestions. It is important to explain the difference between \emph{awareness of coding context} and \emph{adapted to coding context} previously described: in the former case the recommended code is triggered by what has been written in the IDE (\eg it completes the implementation of a method under development), while in the latter the suggested code is adapted (\ie changed) to match up with the current code context (\eg to reuse identifiers present in the code). These two aspects should be combined together to obtain useful recommendations.

Other interesting requirements developers expressed are the awareness of \emph{coding history} and of the \emph{developer's tasks}. \faLightbulbO~Exploiting the change history of the system on which recommendations are generated \cite{robbeshistory,robbes2010improving} can ``\emph{help with repetitive tasks}''. Instead, being aware of the tasks assigned to the developer to which recommendations are proposed can be exploited for a better customization of the recommendations. For example, the recommender can identify the issues assigned to the developer by mining the issue tracker, inferring the one they are working on in the IDE and targeting recommendations aimed at completing the issue being addressed.

\section{Validity Discussion} \label{sec:threats}

Threats to \textbf{construct validity} concern the relationship between theory and observation. The characteristics output of our study are personal opinions of the surveyed developers. To provide information about the ``support'' each characteristic had, we included the number of participants who mentioned the characteristics (at least for top-level categories, complete data available in \cite{replication}).  

Threats to \textbf{internal validity} concern factors internal to our study that could have influenced the results. The participants to our survey are likely more interested in code recommenders than others, thus providing a ``biased view'' of the investigated phenomenon. However, they still provided quite different views on what it is important in code recommenders. 

When presenting the results of our study, we often report ``\emph{quotes}'' from the answers provided by participants. These quotes are not always verbatim, with changes introduced to fix typos or to shorten them without, however, changing their meaning.

To limit subjectivity bias during the open coding procedure, five authors independently inspected each answer we received, with a following discussion aimed at solving conflicts when needed. On top of that, the answers we collected together with the codes (\ie categories of our taxonomy) we assigned them are publicly available in our replication package \cite{replication}. 

Finally, contrary to our expectation, we found difficulties in recruiting software practitioners through MTurk. Indeed, when we run our survey, we did not set the strict selection criteria for participants described in \secref{sec:study1participants}. This resulted in the collection of mostly low-quality answers, often clearly copied/pasted from online sources that we had to exclude. 

Also, we had to forbid the access to our survey from specific countries due to bots providing random answers. At the end, we preferred to collect less answers but of high quality. Indeed, whenever we had a doubt about the quality of answers collected through our survey, we discarded the corresponding response.

Such a ``quality issue'' is less relevant when it comes to answers we collected through our contact network, which represent the majority of completed surveys in our study (49 out of \totalDevs). Indeed, those are all professional developers (\ie working in companies) who voluntarily agreed to participate in our survey. To check the extent to which our taxonomy generalizes to these two different groups of participants we involved (\ie practitioners from MTurk \emph{vs} practitioners in our contact network) we checked the number of categories in our final taxonomy that have been indicated as relevant (i) by both groups, (ii) by MTurk participants only, and (iii) by practitioners in our contact network only. Out of the \characteristics categories in our taxonomy 51 (73\%) have been indicated by participants belonging to both groups, 3 (4\%) by MTurk's participants only, and 16 (23\%) by practitioners in our contact network only. These results indicate an overall ``agreement'' among the two groups for two-thirds of our taxonomy. Also, the fact that only three out of \characteristics categories have been contributed exclusively by MTurk's participants support the validity of the answers collected through this platform, since most of the ``requirements'' we extracted from them have been confirmed by practitioners in our contact network. The three categories being MTurk-only are: \emph{Functionality} $\rightarrow$ \emph{Coverage of support} $\rightarrow$ \emph{Language keywords}, \emph{Functionality} $\rightarrow$ \emph{Recommendation modality} $\rightarrow$ \emph{Open source}, \emph{Functionality} $\rightarrow$ \emph{Recommendation modality} $\rightarrow$ \emph{Proactive: perceives when developer needs help}.

Threats to \textbf{external validity} concern the generalization of our findings. The obvious threat is the limited number of participants involved in the survey (\totalDevs). Such numbers are in line with previously published survey studies (\eg \cite{Forward:DocEng2002,DeSouza:SIGDOC05,Bavota:icse2013}) but, of course, replications can help in corroborating our findings and complementing them.

\section{Related Work} \label{sec:related}

In their seminal paper, Murphy \etal \cite{murphy2006java} found that developers rely on \textit{content assist} (\ie code completion) features in IDEs as much as on other common editing features (\eg copy \& paste).
Such a result showed how improving code completion could have benefited developers. Code recommender systems greatly evolved since then. 

For the sake of brevity, we do not focus on the many works focusing on improving code recommendations (\eg \cite{Xie:msr2006, Moreno:icse2015, Moreno:tse2016, goues:icse2012, Tufano:tosem2019, Li:icse2020, McMillan:tse2013, Wen:icse2021, Nguyen:fse2016, Liu:ase2020, Kim:ASE2009, Allamanis:fse2014, kim2020code, Karampatsis:DLareBest, Watson:icse2020, Tufano:asserts}) but on empirical studies looking at code recommenders from different perspectives.

Proksch \etal \cite{proksch2016evaluating} evaluated a method-call recommender system on a real-world dataset featuring interactions captured in the IDE. They observed that commonly used evaluations based on synthetic datasets extracted \textit{a-posteriori} from released code do not take into account context change: This has a major effect on the prediction quality.

Hellendoorn \etal \cite{hellendoorn2019code} compared code completion models on a real-world dataset and on synthetic datasets. They found that the experimented tools are less accurate on the real-world dataset,  showing that synthetic benchmarks are not representative enough. Moreover, they found that such tools are less accurate in challenging scenarios, when developers would need them the most.

Ciniselli \etal \cite{Ciniselli:msr2022} investigated the extent to which code recommenders copy code from their training set when generating recommendations. They found that $\sim$10\% of of short recommendations represent clones of training set's code. However, as soon as the size of the recommended code increases (\eg a few statements), then it is unlikely that code recommenders copy code from the training set.

M{\u{a}}r{\u{a}}șoiu \etal \cite{muaruasoiu2015empirical} studied how developers use code completion in practice.
They observed that many recommendations are not accepted by the users. Similar findings have been reported by Arrebola and Junior \cite{arrebola2017source}, who advocate for context-awareness. 

Jin and Servant \cite{jin2018hidden} investigated the \textit{hidden costs} of code recommendations. 
They found that the code completion tool they evaluated (IntelliSense) sometimes provides the right recommendation far from the top of the list. They observed that longer lists discourage developers from selecting a recommendation.

Xu \etal \cite{xu2021inide} run a controlled experiment with 31 developers who were asked to complete implementation tasks with and without the support of two code recommenders. They found no significant gain in developers' productivity when using the code recommenders. 

Ziegler \etal \cite{Ziegler:maps2022} run a survey with Copilot users to investigate which quantitative measure better capture their perceived productivity when using the tool. They found that the acceptance rate of shown suggestions is the best predictor of perceived productivity. 

The discussed papers suggest that a lack of grounding in reality may be detrimental for the advancement in such a field. We try to further fill this gap: while previous work mostly focused on specific aspects, such as accuracy \cite{proksch2016evaluating, hellendoorn2019code}, lack of context-awareness \cite{arrebola2017source}, or hidden cost \cite{jin2018hidden}, we provide a complete developer-oriented view on the possible challenges in the design of code recommenders.

\section{Conclusion and Future Work} \label{sec:conclusion}
Our study partially fills the lack of empirical investigations aimed at collecting practitioners' \emph{desiderata} when it comes to code recommenders. We run a survey involving a total of \totalDevs practitioners to investigate characteristics of code recommenders they perceive as important.  As output of our study, we defined a taxonomy of \characteristics characteristics (\figref{fig:results}) that can drive future research in the field. We make all (anonymized) answers we collected, the tags we assigned to them, and study material available in our replication package hosted at \underline{\url{https://code-recommenders.github.io}}.

\eject

Our future works stem from the findings of our surveys, and will focus on improving characteristics of code recommenders that our study highlighted as relevant. We detail three research directions we plan to pursue, with the goal of also showing how our taxonomy can be used to point at future research:

\begin{itemize}

\item \emph{Improving the awareness of code recommenders}. One aspect several of the participants in our study stressed as important is what we defined as the ``awareness of the code recommenders''. In other words, what the information available to the recommender are when it synthesizes suggestions. Being aware of the developers' knowledge (\ie their expertise, past implementation tasks, \etc) could result in more relevant recommendations that might be particularly suitable and easy to understand and reuse for the developer receiving them. Integrating such a ``knowledge'' in the recommender is far from trivial and requires the development of techniques allowing to automatically infer (i) the programming style of software developers, and (ii) their expertises, namely the specific programming languages, libraries, notions (\eg design patterns), \etc they are at ease with. This could be done by mining the past developers' activities from software repositories (\eg versioning system, issue tracker).\smallskip

\item \emph{Making code quality a first-class citizen in code recommendations}. Given the increasing complexity of the recommendations supported by tools such as GitHub Copilot, code quality must become a priority for the recommended code. This is a clear outcome of our survey. Important aspects to focus on are the absence of bugs/vulnerabilities and the promotion of readability and understandability. All these quality aspects can benefit from tailored design decisions made (i) when training the code recommender, by curating the quality of the code snippets composing the training set; and (ii) at post-processing stage, with checks done before triggering the recommendation to the user. Also, alternative recommendations could be ranked based on their quality. \smallskip


\item \emph{Augmenting the coverage of support}. Despite the gigantic steps ahead made in the last few years, code recommenders still struggle in specific coding scenarios for which they do not offer (or offer limited) support. In this context, two interesting research directions stemming from our taxonomy are the (i) better support for multi-lingual code, and (ii) generation of templates rather than raw source code when the recommender is not confident. Concerning the first point, we plan to assess the effectiveness of recent transformer models trained on several programming languages (\eg CodeBERT \cite{feng-etal-2020-codebert}) when dealing with multi-lingual code. Depending on the observed performance, the proposal of alternative strategies to deal with this problem will be investigated. As for the template generation, we plan to work on a model specifically trained for generating abstract code templates rather than raw source code. Such a model could be triggered when the standard code recommender is not confident in suggesting the raw code.

\end{itemize}

\section*{Acknowledgment}
This project has received funding from the European Research Council (ERC) under the European Union's Horizon 2020 research and innovation programme (grant agreement No. 851720).

\balance

\bibliographystyle{IEEEtranS}
\bibliography{main}

\begin{thebibliography}{10}
\providecommand{\url}[1]{#1}
\csname url@samestyle\endcsname
\providecommand{\newblock}{\relax}
\providecommand{\bibinfo}[2]{#2}
\providecommand{\BIBentrySTDinterwordspacing}{\spaceskip=0pt\relax}
\providecommand{\BIBentryALTinterwordstretchfactor}{4}
\providecommand{\BIBentryALTinterwordspacing}{\spaceskip=\fontdimen2\font plus
\BIBentryALTinterwordstretchfactor\fontdimen3\font minus
  \fontdimen4\font\relax}
\providecommand{\BIBforeignlanguage}[2]{{%
\expandafter\ifx\csname l@#1\endcsname\relax
\typeout{** WARNING: IEEEtranS.bst: No hyphenation pattern has been}%
\typeout{** loaded for the language `#1'. Using the pattern for}%
\typeout{** the default language instead.}%
\else
\language=\csname l@#1\endcsname
\fi
#2}}
\providecommand{\BIBdecl}{\relax}
\BIBdecl

\bibitem{mturk}
``Amazon mechanical turk~\url{https://www.mturk.com}.''

\bibitem{copilot}
``Github copilot~\url{https://copilot.github.com}.''

\bibitem{qualtrics}
``Qualtrics~\url{https://www.qualtrics.com}.''

\bibitem{replication}
``Replication package~\url{https://code-recommenders.github.io}.''

\bibitem{Allamanis:fse2014}
M.~Allamanis, E.~T. Barr, C.~Bird, and C.~Sutton, ``Learning natural coding
  conventions,'' in \emph{Proceedings of the 22nd ACM SIGSOFT International
  Symposium on Foundations of Software Engineering}, ser. FSE 2014, 2014, pp.
  281--293.

\bibitem{Amorim:sle2016}
L.~E. d.~S. Amorim, S.~Erdweg, G.~Wachsmuth, and E.~Visser, ``Principled
  syntactic code completion using placeholders,'' ser. SLE 2016, 2016, p.
  163?175.

\bibitem{arrebola2017source}
F.~V. Arrebola and P.~T.~A. Junior, ``On source code completion assistants and
  the need of a context-aware approach,'' in \emph{International Conference on
  Human Interface and the Management of Information}.\hskip 1em plus 0.5em
  minus 0.4em\relax Springer, 2017, pp. 191--201.

\bibitem{Bavota:icse2013}
G.~Bavota, B.~Dit, R.~Oliveto, M.~D. Penta, D.~Poshyvanyk, and A.~D. Lucia,
  ``An empirical study on the developers' perception of software coupling,'' in
  \emph{35th International Conference on Software Engineering, {ICSE} '13, San
  Francisco, CA, USA, May 18-26, 2013}.\hskip 1em plus 0.5em minus 0.4em\relax
  {IEEE} Computer Society, 2013, pp. 692--701.

\bibitem{Bavota:emse2014}
G.~Bavota, A.~D. Lucia, A.~Marcus, and R.~Oliveto, ``Automating extract class
  refactoring: an improved method and its evaluation,'' \emph{Empir. Softw.
  Eng.}, vol.~19, no.~6, pp. 1617--1664, 2014.

\bibitem{Buse:tse2010}
R.~P.~L. Buse and W.~Weimer, ``Learning a metric for code readability,''
  \emph{IEEE Transactions on Software Engineering}, vol.~36, no.~4, pp.
  546--558, 2010.

\bibitem{ciniselli:tse2021}
M.~Ciniselli, N.~Cooper, L.~Pascarella, A.~Mastropaolo, E.~Aghajani,
  D.~Poshyvanyk, M.~D. Penta, and G.~Bavota, ``An empirical study on the usage
  of transformer models for code completion,'' \emph{IEEE Transactions on
  Software Engineering}, no.~01, pp. 1--1, 2022.

\bibitem{Ciniselli:msr2022}
M.~Ciniselli, L.~Pascarella, and G.~Bavota, ``To what extent do deep
  learning-based code recommenders generate predictions by cloning code from
  the training set?'' in \emph{{IEEE/ACM} 19th International Conference on
  Mining Software Repositories, {MSR} 2022, Pittsburgh, PA, USA, May 23-24,
  2022}.\hskip 1em plus 0.5em minus 0.4em\relax {IEEE}, 2022, pp. 167--178.

\bibitem{DeSouza:SIGDOC05}
S.~C.~B. de~Souza, N.~Anquetil, and K.~M. de~Oliveira, ``A study of the
  documentation essential to software maintenance,'' in \emph{Proceedings of
  the 23rd Annual International Conference on Design of Communication:
  Documenting \& Designing for Pervasive Information}, ser. SIGDOC '05.\hskip
  1em plus 0.5em minus 0.4em\relax ACM, 2005, pp. 68--75.

\bibitem{dorn2012general}
J.~Dorn, ``A general software readability model,'' \emph{MCS Thesis available
  from (http://www. cs. virginia. edu/weimer/students/dorn-mcs-paper. pdf)},
  vol.~5, pp. 11--14, 2012.

\bibitem{feng-etal-2020-codebert}
Z.~Feng, D.~Guo, D.~Tang, N.~Duan, X.~Feng, M.~Gong, L.~Shou, B.~Qin, T.~Liu,
  D.~Jiang, and M.~Zhou, ``{C}ode{BERT}: A pre-trained model for programming
  and natural languages,'' in \emph{Findings of the Association for
  Computational Linguistics: EMNLP 2020}.\hskip 1em plus 0.5em minus
  0.4em\relax Online: Association for Computational Linguistics, Nov. 2020, pp.
  1536--1547.

\bibitem{Forward:DocEng2002}
A.~Forward and T.~C. Lethbridge, ``The relevance of software documentation,
  tools and technologies: A survey,'' in \emph{Proc. of the 2002 ACM Symp. on
  Doc. Eng. (DocEng)}.\hskip 1em plus 0.5em minus 0.4em\relax ACM, 2002, pp.
  26--33.

\bibitem{Foster:icse2012}
S.~R. Foster, W.~G. Griswold, and S.~Lerner, ``Witchdoctor: Ide support for
  real-time auto-completion of refactorings,'' in \emph{2012 34th International
  Conference on Software Engineering (ICSE)}, 2012, pp. 222--232.

\bibitem{hellendoorn2019code}
V.~J. Hellendoorn, S.~Proksch, H.~C. Gall, and A.~Bacchelli, ``When code
  completion fails: A case study on real-world completions,'' in \emph{2019
  IEEE/ACM 41st International Conference on Software Engineering (ICSE)}.\hskip
  1em plus 0.5em minus 0.4em\relax IEEE, 2019, pp. 960--970.

\bibitem{Xing:icpc2018}
X.~Hu, G.~Li, X.~Xia, D.~Lo, and Z.~Jin, ``Deep code comment generation,'' ser.
  ICPC '18, 2018.

\bibitem{huang2020towards}
Y.~Huang, S.~Huang, H.~Chen, X.~Chen, Z.~Zheng, X.~Luo, N.~Jia, X.~Hu, and
  X.~Zhou, ``Towards automatically generating block comments for code
  snippets,'' \emph{Information and Software Technology}, vol. 127, p. 106373,
  2020.

\bibitem{jin2018hidden}
X.~Jin and F.~Servant, ``The hidden cost of code completion: Understanding the
  impact of the recommendation-list length on its efficiency,'' in
  \emph{Proceedings of the 15th International Conference on Mining Software
  Repositories}, 2018, pp. 70--73.

\bibitem{Karampatsis:DLareBest}
\BIBentryALTinterwordspacing
R.~Karampatsis and C.~A. Sutton, ``Maybe deep neural networks are the best
  choice for modeling source code,'' \emph{CoRR}, vol. abs/1903.05734, 2019.
  [Online]. Available: \url{http://arxiv.org/abs/1903.05734}
\BIBentrySTDinterwordspacing

\bibitem{Kim:ASE2009}
J.~{Kim}, S.~{Lee}, S.~{Hwang}, and S.~{Kim}, ``Adding examples into java
  documents,'' in \emph{2009 IEEE/ACM International Conference on Automated
  Software Engineering}, 2009, pp. 540--544.

\bibitem{kim2020code}
S.~Kim, J.~Zhao, Y.~Tian, and S.~Chandra, ``Code prediction by feeding trees to
  transformers,'' \emph{arXiv preprint arXiv:2003.13848}, 2020.

\bibitem{kyaw2018proposal}
H.~H.~S. Kyaw, S.~T. Aung, H.~A. Thant, and N.~Funabiki, ``A proposal of code
  completion problem for java programming learning assistant system,'' in
  \emph{Conference on Complex, Intelligent, and Software Intensive
  Systems}.\hskip 1em plus 0.5em minus 0.4em\relax Springer, 2018, pp.
  855--864.

\bibitem{goues:icse2012}
C.~Le~Goues, M.~Dewey-Vogt, S.~Forrest, and W.~Weimer, ``A systematic study of
  automated program repair: Fixing 55 out of 105 bugs for \$8 each,'' in
  \emph{2012 34th International Conference on Software Engineering (ICSE)},
  2012, pp. 3--13.

\bibitem{Li:icse2020}
Y.~Li, S.~Wang, and T.~N. Nguyen, ``Dlfix: Context-based code transformation
  learning for automated program repair,'' in \emph{Proceedings of the ACM/IEEE
  42nd International Conference on Software Engineering}, ser. ICSE '20, 2020,
  p. 602?614.

\bibitem{Liu:ase2020}
F.~Liu, G.~Li, Y.~Zhao, and Z.~Jin, ``Multi-task learning based pre-trained
  language model for code completion,'' in \emph{Proceedings of the 35th
  IEEE/ACM International Conference on Automated Software Engineering}, ser.
  ASE 2020.\hskip 1em plus 0.5em minus 0.4em\relax Association for Computing
  Machinery, 2020.

\bibitem{muaruasoiu2015empirical}
M.~M{\u{a}}r{\u{a}}șoiu, L.~Church, and A.~Blackwell, ``An empirical
  investigation of code completion usage by professional software developers,''
  in \emph{Proceedings of the 26th Annual Workshop of the Psychology of
  Programming Interest Group}, 2015.

\bibitem{marri2009improving}
M.~R. Marri, S.~Thummalapenta, and T.~Xie, ``Improving software quality via
  code searching and mining,'' in \emph{2009 ICSE Workshop on Search-Driven
  Development-Users, Infrastructure, Tools and Evaluation}.\hskip 1em plus
  0.5em minus 0.4em\relax IEEE, 2009, pp. 33--36.

\bibitem{McMillan:tse2013}
C.~McMillan, D.~Poshyvanyk, M.~Grechanik, Q.~Xie, and C.~Fu, ``Portfolio:
  Searching for relevant functions and their usages in millions of lines of
  code,'' \emph{{ACM} Trans. Softw. Eng. Methodol.}, vol.~22, no.~4, pp.
  37:1--37:30, 2013.

\bibitem{mi2018improving}
Q.~Mi, J.~Keung, Y.~Xiao, S.~Mensah, and Y.~Gao, ``Improving code readability
  classification using convolutional neural networks,'' \emph{Information and
  Software Technology}, vol. 104, pp. 60--71, 2018.

\bibitem{Moreno:icse2015}
L.~Moreno, G.~Bavota, M.~Di~Penta, R.~Oliveto, and A.~Marcus, ``How can i use
  this method?'' in \emph{Proceedings of the 37th International Conference on
  Software Engineering - Volume 1}, ser. ICSE '15, 2015, p. 880?890.

\bibitem{Moreno:tse2016}
L.~Moreno, G.~Bavota, M.~D. Penta, R.~Oliveto, A.~Marcus, and G.~Canfora,
  ``Arena: An approach for the automated generation of release notes,''
  \emph{IEEE Transactions on Software Engineering}, vol.~43, no.~2, pp.
  106--127, 2017.

\bibitem{murphy2006java}
G.~C. Murphy, M.~Kersten, and L.~Findlater, ``How are java software developers
  using the elipse ide?'' \emph{IEEE software}, vol.~23, no.~4, pp. 76--83,
  2006.

\bibitem{Nasehi:icsm2012}
S.~M. {Nasehi}, J.~{Sillito}, F.~{Maurer}, and C.~{Burns}, ``What makes a good
  code example?: A study of programming q a in stackoverflow,'' in \emph{2012
  28th IEEE International Conference on Software Maintenance (ICSM)}, 2012, pp.
  25--34.

\bibitem{Nguyen:fse2016}
T.~Nguyen, P.~C. Rigby, A.~T. Nguyen, M.~Karanfil, and T.~N. Nguyen, ``T2api:
  Synthesizing api code usage templates from english texts with statistical
  translation,'' in \emph{Proceedings of the 2016 24th ACM SIGSOFT
  International Symposium on Foundations of Software Engineering}, ser. FSE
  2016, 2016, p. 1013?1017.

\bibitem{Perelman:pldi2012}
D.~Perelman, S.~Gulwani, T.~Ball, and D.~Grossman, ``Type-directed completion
  of partial expressions,'' in \emph{Proceedings of the 33rd ACM SIGPLAN
  Conference on Programming Language Design and Implementation}, ser. PLDI '12,
  2012, p. 275?286.

\bibitem{posnett2011simpler}
D.~Posnett, A.~Hindle, and P.~Devanbu, ``A simpler model of software
  readability,'' in \emph{Proceedings of the 8th working conference on mining
  software repositories}, 2011, pp. 73--82.

\bibitem{proksch2016evaluating}
S.~Proksch, S.~Amann, S.~Nadi, and M.~Mezini, ``Evaluating the evaluations of
  code recommender systems: a reality check,'' in \emph{2016 31st IEEE/ACM
  International Conference on Automated Software Engineering (ASE)}.\hskip 1em
  plus 0.5em minus 0.4em\relax IEEE, 2016, pp. 111--121.

\bibitem{Reiss:ase2007}
S.~P. Reiss, ``Automatic code stylizing,'' in \emph{Proceedings of the
  Twenty-Second IEEE/ACM International Conference on Automated Software
  Engineering}, ser. ASE '07, 2007, p. 74?83.

\bibitem{robbeshistory}
R.~Robbes and M.~Lanza, ``How program history can improve code completion,'' in
  \emph{2008 23rd IEEE/ACM International Conference on Automated Software
  Engineering}, 2008, pp. 317--326.

\bibitem{robbes2010improving}
------, ``Improving code completion with program history,'' \emph{Automated
  Software Engineering}, vol.~17, no.~2, pp. 181--212, 2010.

\bibitem{robillard:recommenders}
M.~P. Robillard, W.~Maalej, R.~J. Walker, and T.~Zimmermann,
  \emph{Recommendation Systems in Software Engineering}.\hskip 1em plus 0.5em
  minus 0.4em\relax Springer Publishing Company, Incorporated, 2014.

\bibitem{scalabrino2018comprehensive}
S.~Scalabrino, M.~Linares-V{\'a}squez, R.~Oliveto, and D.~Poshyvanyk, ``A
  comprehensive model for code readability,'' \emph{Journal of Software:
  Evolution and Process}, vol.~30, no.~6, p. e1958, 2018.

\bibitem{schuster2020autocomplete}
R.~Schuster, C.~Song, E.~Tromer, and V.~Shmatikov, ``You autocomplete me:
  Poisoning vulnerabilities in neural code completion,'' 2020.

\bibitem{svyatkovskiy2020intellicode}
A.~Svyatkovskiy, S.~K. Deng, S.~Fu, and N.~Sundaresan, ``Intellicode compose:
  Code generation using transformer,'' \emph{arXiv preprint arXiv:2005.08025},
  2020.

\bibitem{svyatkovskiy2020fast}
A.~Svyatkovskiy, S.~Lee, A.~Hadjitofi, M.~Riechert, J.~V. Franco, and
  M.~Allamanis, ``Fast and memory-efficient neural code completion,'' in
  \emph{18th {IEEE/ACM} International Conference on Mining Software
  Repositories, {MSR} 2021}.\hskip 1em plus 0.5em minus 0.4em\relax {IEEE},
  2021, pp. 329--340.

\bibitem{Tamrawi:2011}
A.~Tamrawi, T.~T. Nguyen, J.~M. Al-Kofahi, and T.~N. Nguyen, ``Fuzzy set and
  cache-based approach for bug triaging,'' in \emph{Proceedings of the 19th ACM
  SIGSOFT Symposium and the 13th European Conference on Foundations of Software
  Engineering}, ser. ESEC/FSE '11, 2011, p. 365?375.

\bibitem{Tsantalis:saner2018}
N.~Tsantalis, T.~Chaikalis, and A.~Chatzigeorgiou, ``Ten years of jdeodorant:
  Lessons learned from the hunt for smells,'' in \emph{25th International
  Conference on Software Analysis, Evolution and Reengineering, {SANER} 2018},
  R.~Oliveto, M.~D. Penta, and D.~C. Shepherd, Eds.\hskip 1em plus 0.5em minus
  0.4em\relax {IEEE} Computer Society, 2018, pp. 4--14.

\bibitem{Tufano:asserts}
M.~Tufano, D.~Drain, A.~Svyatkovskiy, and N.~Sundaresan, ``Generating accurate
  assert statements for unit test cases using pretrained transformers,''
  \emph{CoRR}, vol. abs/2009.05634, 2020.

\bibitem{Tufano:tosem2019}
M.~Tufano, C.~Watson, G.~Bavota, M.~{Di Penta}, M.~White, and D.~Poshyvanyk,
  ``An empirical study on learning bug-fixing patches in the wild via neural
  machine translation,'' \emph{{ACM} Trans. Softw. Eng. Methodol.}, vol.~28,
  no.~4, pp. 19:1--19:29, 2019.

\bibitem{Tufano:icse2022}
R.~Tufano, S.~Masiero, A.~Mastropaolo, L.~Pascarella, D.~Poshyvanyk, and
  G.~Bavota, ``Using pre-trained models to boost code review automation,'' in
  \emph{44th {IEEE/ACM} 44th International Conference on Software Engineering,
  {ICSE} 2022, Pittsburgh, PA, USA, May 25-27, 2022}.\hskip 1em plus 0.5em
  minus 0.4em\relax {IEEE}, 2022, pp. 2291--2302.

\bibitem{Tufano:icse2021}
R.~Tufano, L.~Pascarella, M.~Tufano, D.~Poshyvanyk, and G.~Bavota, ``Towards
  automating code review activities,'' in \emph{43rd {IEEE/ACM} International
  Conference on Software Engineering, {ICSE} 2021, Madrid, Spain, 22-30 May
  2021}.\hskip 1em plus 0.5em minus 0.4em\relax {IEEE}, 2021, pp. 163--174.

\bibitem{wang2020towards}
W.~Wang, S.~Shen, G.~Li, and Z.~Jin, ``Towards full-line code completion with
  neural language models,'' \emph{arXiv preprint arXiv:2009.08603}, 2020.

\bibitem{Watson:icse2020}
C.~Watson, M.~Tufano, K.~Moran, G.~Bavota, and D.~Poshyvanyk, ``On learning
  meaningful assert statements for unit test cases,'' in \emph{Proceedings of
  the 42nd International Conference on Software Engineering, {ICSE} 2020},
  2020, p. To Appear.

\bibitem{Wen:icse2021}
F.~Wen, E.~Aghajani, C.~Nagy, M.~Lanza, and G.~Bavota, ``Siri, write the next
  method,'' in \emph{43rd {IEEE/ACM} International Conference on Software
  Engineering, {ICSE} 2021, Madrid, Spain, 22-30 May 2021}.\hskip 1em plus
  0.5em minus 0.4em\relax {IEEE}, 2021, pp. 138--149.

\bibitem{Xia:tse2016}
X.~Xia, D.~Lo, Y.~Ding, J.~M. Al-Kofahi, T.~N. Nguyen, and X.~Wang, ``Improving
  automated bug triaging with specialized topic model,'' \emph{IEEE
  Transactions on Software Engineering}, vol.~43, no.~3, pp. 272--297, 2017.

\bibitem{Xie:msr2006}
T.~Xie and J.~Pei, ``Mapo: Mining api usages from open source repositories,''
  ser. MSR '06, 2006.

\bibitem{xu-etal-2020-incorporating}
F.~F. Xu, Z.~Jiang, P.~Yin, B.~Vasilescu, and G.~Neubig, ``Incorporating
  external knowledge through pre-training for natural language to code
  generation,'' in \emph{Proceedings of the 58th Annual Meeting of the
  Association for Computational Linguistics}.\hskip 1em plus 0.5em minus
  0.4em\relax Association for Computational Linguistics, Jul. 2020.

\bibitem{xu2021inide}
F.~F. Xu, B.~Vasilescu, and G.~Neubig, ``In-ide code generation from natural
  language: Promise and challenges,'' 2021.

\bibitem{Ziegler:maps2022}
A.~Ziegler, E.~Kalliamvakou, X.~A. Li, A.~Rice, D.~Rifkin, S.~Simister,
  G.~Sittampalam, and E.~Aftandilian, ``Productivity assessment of neural code
  completion,'' ser. MAPS 2022, 2022, p. 21?29.

\end{thebibliography}
\newpage

\end{document}